\documentclass[aps,prc,preprint,superscriptaddress]{revtex4-1}
\usepackage{graphicx}

\begin{document}

\title{The Influence of in-medium NN cross-sections, symmetry potential and impact parameter on the isospin observables}


\author{Yingxun Zhang} 
\affiliation{China Institute of Atomic Energy, P.O. Box 275 (10), Beijing 102413, P.R. China}
\affiliation{Joint Institute of Nuclear Astrophysics, Michigan State University, E. Lansing, MI 48824, USA}
\author{D.D.S.Coupland}
\affiliation{National Superconducting Cyclotron Laboratory and Physics and Astronomy Department, Michigan State University, East Lansing, MI 48824, USA}
\author{P.Danielewicz}
\affiliation{Joint Institute of Nuclear Astrophysics, Michigan State University, E. Lansing, MI 48824, USA}
\affiliation{National Superconducting Cyclotron Laboratory and Physics and Astronomy Department, Michigan State University, East Lansing, MI 48824, USA}
\author{Zhuxia Li} 
\affiliation{China Institute of Atomic Energy, P.O. Box 275 (10), Beijing 102413, P.R. China}

\author{Hang Liu} 
\affiliation{Texas Advanced Computing Center, University of Texas, Austin, TX 78758, USA}

\author{Fei Lu} 
\affiliation{Joint Institute of Nuclear Astrophysics, Michigan State University, E. Lansing, MI 48824, USA}
\affiliation{Department of Physics and State Key Laboratory of Nuclear Physics and Technology, Peking University, Beijing 100871, P.R. China}

\author{W.G.Lynch} 
\affiliation{Joint Institute of Nuclear Astrophysics, Michigan State University, E. Lansing, MI 48824, USA}
\affiliation{National Superconducting Cyclotron Laboratory and Physics and Astronomy Department, Michigan State University, East Lansing, MI 48824, USA}

\author{M.B.Tsang} 
\affiliation{Joint Institute of Nuclear Astrophysics, Michigan State University, E. Lansing, MI 48824, USA}
\affiliation{National Superconducting Cyclotron Laboratory and Physics and Astronomy Department, Michigan State University, East Lansing, MI 48824, USA}


\date{\today}

\begin{abstract}
We explore the influence of in-medium nucleon-nucleon cross section, symmetry potential and impact parameter on isospin sensitive observables in intermediate-energy heavy-ion collisions with the ImQMD05 code, a modified version of Quantum Molecular Dynamics model. At incident velocities above the Fermi velocity, we  find that the density dependence of symmetry potential plays a more important role on the double neutron to proton ratio $DR(n/p)$ and the isospin transport ratio $R_i$ than the in-medium nucleon-nucleon cross sections, provided that the latter are constrained to a fixed total NN collision rate. We also explore both $DR(n/p)$ and $R_i$ as a function of the impact parameter. Since the copious production of intermediate mass fragments is a distinguishing feature of intermediate-energy heavy-ion collisions, we examine the isospin transport ratios constructed from different groups of fragments. We find that the values of the isospin transport ratios for projectile rapidity fragments with $Z\ge20$ are greater than those constructed from the entire projectile rapidity source. We believe experimental investigations of this phenomenon can be performed. These may provide significant tests of fragmentation time scales predicted by ImQMD calculations.
\end{abstract}


\maketitle

\section{Introduction}
The nuclear symmetry energy plays an important role in the properties of nuclei and neutron stars~\cite{Latti2001,Latti2004,Stein2005,Anna2006,Yakov2004}. To a good approximantion, it can be written as
\begin{equation}
E_{sym}=S(\rho)\delta^2.
\end{equation}
where $\delta=(\rho_n-\rho_p)/(\rho_n+\rho_p)$, is the isospin asymmetry; $\rho_{n}$, $\rho_{p}$, are the neutron, proton densities, and $S(\rho)$ describes the density dependence of the symmetry energy. Theoretical predictions for $S(\rho)$ from microscopic nucleon-nucleon interactions show large uncertainties, especially in the region of suprasaturation density \cite{Brown91,BALi08}. Constraining the density dependence of the symmetry energy has become one of the main goals in nuclear physics and has stimulated many theoretical and experimental studies \cite{BALi08,Danie02,Fuch06,Garg04,HSXu00,Tsang01,Shett04,Tsang04,LWCh04,qfli05,qfli06,TXLiu07,BALi05,Fami06,BALi97,BALi06,BALi00,BALi04,zhang05,zhang08,Yong06,Tsang09,Gior10,Napo10}.
Heavy Ion Collisions (HIC) with asymmetric nuclei provide a unique opportunity for laboratory studies of the density dependence of the symmetry energy because a large range of densities can be momentarily achieved during HICs. The experimental strategy is to
measure the collisions of neutron-rich and neutron-poor system and compare them in order to create observable that are primarily sensitive to the density dependence of the symmetry energy. Comparisons of these observables to theoretical predictions provide the sought-after constraints.

Several semi-classical transport theoretical models have been developed to to simulate the nucleus-nucleus collisions and have been used to obtain constraints on the symmetry energy. In Section II. A, we discuss and compare the approaches using the Boltzmann-Uehling-Uhlenbeck (BUU) equation and the Molecular Dynamics Model (QMD). In this work, we choose to simulate nuclear collisions with the code ImQMD05 developed at the Chinese Institute of Atomic Energy (CIAE) \cite{zhang05,zhang06,zhang07}. Details of this code are described in section II. B.  In Section II. C, we show the dynamics of the simulated collisions of $^{112,124}Sn +^{112,124}Sn$ at E/A = 50MeV. The influences of the density dependence of symmetry energy, in-medium nucleon-nucleon cross sections, and impact parameters on isospin sensitive observables for the simulated reactions are investigated in Section III. We focus our studies mainly on two experimental observables.  We examine the ratio between neutron and proton yields in Section III.A, and the isospin transport ratio in Section III.B where we also discuss the influence of cluster emission on isospin transport ratio. Since the main goal of this paper is to investigate the influences of various parameters used as input for the transport model there is no attempt to obtain the best fit to the data as was done in Ref.~\cite{Tsang09}.  We summarize our findings in section IV.

\section{Transport description of Heavy Ion Collisions}

One frequently utilized transport models to describe the heavy ion collisions is the Boltzmann-Uehling-Uhlenbeck (BUU) equation, which provides an approximate Wigner transform of the one-body density matrix as its solution\cite{Bertsch88}. Another frequently utilized approaches, known as the Molecular Dynamics Model (QMD) represent the individual nucleons as Gaussian "wave-packet" with mean values that move in according the Ehrenfest theorem; i.e. Hamilton's equations\cite{Aiche87}. Even though we do not use the BUU equations in our work, it is useful to discuss the similarities and differences between the two approaches.

\subsection{Remarks on QMD and BUU}
At the code level, both BUU and QMD models propagate particles classically under the influence of a mean field potential, which is calculated self-consistently the positions and momenta of the particles, and allow scattering by nucleon-nucleon collisions due to the residual interaction. The Pauli principle in both approaches is enforced by application of Pauli blocking factors. These similarities in implementation have lead to similarities in predictions for many collision observables \cite{Aich89}.

There are also significant differences in these approaches. In the BUU equations, each nucleon is represented by 200-1000 test particels that generate the mean field and suffer the collisions. In QMD, there is one test particle per nucleon. A-body correlations and cluster formation are not native to the original BUU approach; which is supposed to provide the Wigner transform of the one body density matrix. On the other hand, many-body correlations and fluctuations can arise from the A-body dynamics of QMD approach. Such A-body correlations are suppressed in BUU approach, but correlations can arise in both approaches from the amplification of mean field instabilities in spinodal region \cite{Chomaz04}. Collision algorithms in the QMD approach modify the momenta of individual nucleons, while in BUU approach, only the momenta of test particles are modified. Depending on the details of the in-medium cross sections that are implemented, the blocking of collisions can also be more restrictive for QMD than for BUU, leading to fewer collisions and therefore a greater transparency. Since there are typically more than 100 test particles per nucleon, collision induced fluctuations are smaller in BUU than in QMD possibly suppressing the fragment formation rates.

Fragments can be formed in QMD approaches due to the A-body correlations and these correlations are mapped onto the asymptotic final fragments by a spanning tree algorithm. Serval different methods have been developed to allow BUU codes to calculate cluster production. In the Stochastic Mean Field (SMF) approach, for example, the time evolution of the one-body phase-space distribution $f$ is governed by the nuclear mean-field, two-body scattering, and a fluctuating (stochastic) term, which causes the fragmentation\cite{Baran02,Colonna98,Rizzo08,Baran04}. Such fluctuations are sometimes achieved by reducing the number of test particles in a BUU simulation to about 50 per nucleon. The pBUU code developed in Ref.\cite{Dan00} provides a sophisticated theoretical description of clusters up to A=3, but lacks heavier particles, including alphas. Also, the coalescence approximation or the spanning tree method has predicted light cluster yields from the sampled test particle distributions. Of these three methods, the production of clusters has a feedback on the collision dynamics in the first two methods, while the coalescence or the spanning tree methods do not.

Without modifications to produce clusters, direct comparisons of BUU calculations to the observables for Intermediate Mass Fragments (IMF's), such as the rapidity dependence of isospin transport ratios extracted from the yields of mirror nuclear fragments, are rarely performed. Such fragment observables have been modeled within Stochastic Mean Fields (SMF) models, which add fluctuations to the BUU and within the QMD approach.

There are many versions of BUU and QMD codes used by different groups. Even though there is substantial overlap between the results, the extracted constraints on the symmetry energy based on isospin diffusion data obtained from QMD approach (ImQMD05\cite{zhang05,zhang06,zhang07,zhang08}), tend to favor a somewhat softer density dependence of symmetry energy term in the equation of state than those currently provided using BUU approaches, such as the IBUU04\cite{BALi05,LWChen05}, SMF\cite{Baran02,Colonna98}, and pBUU\cite{Dan00} codes.

In some case, the BUU approaches predict much smaller values for the double n/p ratios, $DR(n/p)$, than the experimental results. It is important to determine to what extent these differences arise from (1) differences in the underlying transport models like those discussed in the preceding paragraphs, or from (2) differences in the mean fields or in medium cross sections used in these calculations or (3) in their numerical implementation in the codes.

Differences in the results between the various QMD and BUU approaches can arise from the transport parameters used in the codes. The two main ingredients in transport models, the nucleonic mean fields and nucleon-nucleon binary scattering cross sections, are not treated consistently with the same microscopic nucleon-nucleon interactions used for both mean field and collisions. Thus, there can be differences in the approximations made to adapt these quantities to the calculations and associated differences in the predictions made with them. While the effects of the symmetry energy around saturation density and the isospin dependence of in-medium nucleon-nucleon cross sections have been studied in several BUU codes\cite{BALi05}, it has not been similarly studied in QMD models. This paper provides a constrained exploration on the sensitivity to the isospin dependence of in-medium corrections by expanding upon the ImQMD05 calculations published in Ref. \cite{Tsang09}.

\subsection{Brief description of ImQMD05}
Early implementations of the QMD model did not explore the density dependence of nuclear symmetry potential\cite{Aich91}. At CIAE, we developed and successfully applied a new QMD code, labeled ImQMD, to study heavy ion reactions close to the Coulomb barrier. Such as the excitation function for fusion cross sections\cite{Wang02}. In this code, we incorporated a symmetry energy potential that depended linearly on density. Modifications incorporated in the later ImQMD05 version of this code include mean field potentials calculated using a Skyrme energy density functional with options for different forms of the density dependence of the symmetry potential\cite{zhang05,zhang06,zhang07}. With these modifications, ImQMD05 has successfully described the multiplicity of reaction products, collective flows and stopping powers in intermediate energy Heavy Ion Collisions (HICs). More recently, it has been used to constrain the density dependence of symmetry energy at sub-saturation density using three experimental observables: (1) double n/p ratios, (2) isospin diffusion constructed with isoscaling parameters, and (3) with the ratio of mirror nuclei as a function of rapidity for $^{112,124}Sn+^{112,124}Sn$ at E/A=50MeV\cite{Tsang09}.

In this section, we will describe in more detail the theoretical formalisms used in the ImQMD05 code\cite{zhang08,zhang05,zhang06,zhang07}. Within this code, the nucleonic mean fields acting on nucleon wavepackets are derived from a potential energy density functional where the potential energy $U$ includes the full Skyrme potential energy with the spin-orbit term omitted:
\begin{equation}
U=U_{\rho}+U_{md}+U_{coul} \label{upot}
\end{equation}
Here, $U_{coul}$ is the Coulomb energy, while the nuclear contributions can be represented in local form with
\begin{equation}
U_{\rho,md}=\int u_{\rho,md} d^{3}r \label{urhomd}
\end{equation}  						
and,
\begin{eqnarray}
u_{\rho}=&&\frac{\alpha}{2}\frac{\rho^{2}}{\rho_{0}}+\frac{\beta}{\eta+1}\frac{\rho^{\eta+1}}{\rho^{\eta}_{0}}+\frac{g_{sur}}{2\rho_{0}}(\nabla \rho)^2 \nonumber\\
&&+\frac{g_{sur,iso}}{\rho_{0}}(\nabla (\rho_{n}-\rho_{p}))^2 \nonumber\\
&&+\frac{C_{s}}{2}(\frac{\rho}{\rho_{0}})^{\gamma_i}\delta^{2}\rho+g_{\rho\tau}\frac{\rho^{8/3}}{\rho_{0}^{5/3}} \label{urho}
\end{eqnarray}
where the asymmetry $\delta=(\rho_n-\rho_p)/(\rho_n+\rho_p)$, and $\rho_{n}$, $\rho_{p}$ are the neutron and proton densities. In the present work, a symmetry potential energy density of the form $\frac{C_{s}}{2}(\frac{\rho}{\rho_{0}})^{\gamma_i}\delta^{2}\rho$ is used in transport model comparisons. The energy density associated with the mean-field momentum dependence is represented by
\begin{eqnarray}
u_{md}=\frac{1}{2\rho_{0}}\sum_{N_1,N_2} \frac{1}{16\pi^{6}}\int d^{3}p_{1}d^{3}p_{2}f_{N_{1}}(\vec{p}_1)f_{N_{2}}(\vec{p}_2)\nonumber\\
1.57[\ln(1+5\times 10^{-4}(\Delta p)^2)]^2\;. \label{eq:umd}
\end{eqnarray}
Here, $f_{N}$ are nucleon Wigner functions, $\Delta p=|\vec{p}_1-\vec{p}_2 |$ , and the energy is given in $MeV$ and the momenta in $MeV/c$.
The resulting interaction between wavepackets in Eq. (\ref{eq:umd}) is described in Ref. \cite{Aiche87}. All of the calculations use the potential given above, and the coefficients used in formula (\ref{urho}) are directly related to the standard Skyrme interaction parameters, with $\alpha=-356 MeV$, $\beta =303 MeV$, $\eta =7/6$, $g_{sur}=19.47 MeVfm^2$, $g_{sur,iso} =-11.35 MeVfm^{2}$, $C_{s}=35.19 MeV$, and $g_{\rho\tau} =0$. Eq. (\ref{eq:umd}) modifies the EoS so that it no longer matches the EoS obtained with
the Skyrme interaction parameters at T=0MeV. Appendix discuss the influence on isospin observabels when the isoscalar part of the EOS obtained with Skyrme interaction is changed by the introduction of momentum dependence to Eq. (\ref{eq:umd}). Within statistical error, we found that it has no significant effect on the isospin observables studied in this paper.

The isospin-dependent in-medium nucleon-nucleon scattering cross sections in the collision term are assumed to be the form: $\sigma^{*}_{nn/np}=(1-\xi(E_{beam})\rho/\rho_{0})\sigma^{free}_{nn/np}$, where $\xi(E_{beam})=0.2$ for $E_{beam}=50AMeV$, and the isospin dependent $\sigma^{free}_{nn/np}$ is taken from Ref. \cite{Cugn96}. The in-medium isospin dependent nucleon-nucleon (NN) differential cross section in free space is adopted from Ref.~\cite{Cugn96},
and isospin dependent Pauli blocking effects are the same as in \cite{zhang05,zhang06,zhang07}. Asymptotic clusters are constructed by means of the minimum tree spanning method widely used in QMD calculations, in which particles with relative momentum smaller than $P_{0}$ and relative distance smaller than $R_{0}$ coalesce into one cluster. In the present work, the values of $R_{0}=3.5fm$ and $P_{0}=250MeV/c$ are employed.

From the adopted interaction in the ImQMD05, we construct the density dependence of symmetry energy for cold nuclear matter as follows:
\begin{equation}
S(\rho)=\frac{1}{3}\frac{\hbar^2}{2m}\rho^{2/3}_{0}(\frac{3\pi^2}{2}\frac{\rho}{\rho_{0}})^{2/3}+\frac{C_{s}}{2}(\frac{\rho}{\rho_{0}})^{\gamma_{i}},\label{srho}
\end{equation}
where m is the nucleon mass and the symmetry coefficient $C_s=35.19MeV$. Using this particular parameterization, the symmetry energy at subsaturation densities increases with decreasing  $\gamma_i$, while the opposite is true for supranormal densities. In general, the EoS is labeled as stiff-asy for $\gamma_i>1$, and as soft-asy for $\gamma_i<1$. In the present work, calculations are performed using $\gamma_i=0.5$ and $2.0$ as representative cases of a soft asy-EOS and a stiff asy-EoS, respectively. For illustration, the density dependencies of symmetry energy for $\gamma_i=0.5$(solid line) and $\gamma_i=2.0$(dashed line) are plotted in Fig. 1. For this work, the region of interest is mainly at subsaturation densities, $\rho\le0.16fm^{-3}$.
 \begin{figure}[htbp]
     \centering
 \includegraphics[angle=90,width=0.5\textwidth]{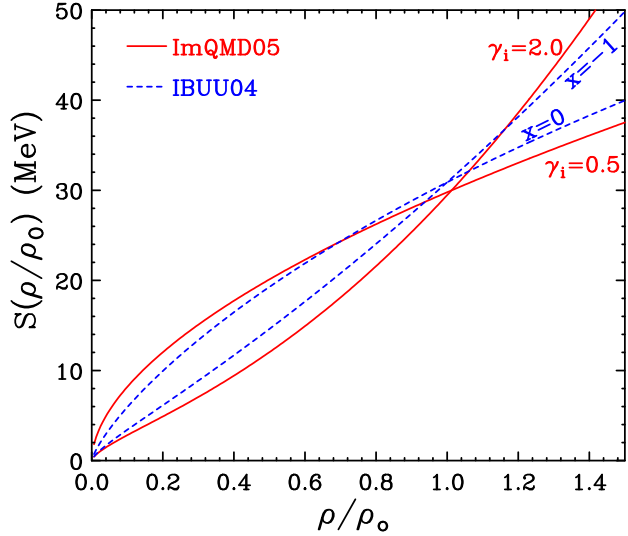} 
 \setlength{\abovecaptionskip}{40pt}
 \caption{(Color online) The symmetry energy for cold nuclear matter plotted as a function of density for $\gamma_i=0.5$ and $2.0$ (solid line) for symmetry potentials used in ImQMD05 simulations (Eq. \ref{srho}). For reference, we show the symmetry energies used in IBUU04, for x=0 and -1 (dot dashed lines) from Ref[29] .\label{fig1}}
  \setlength{\belowcaptionskip}{40pt}
 \end{figure}

\subsection{Reaction Dynamics at E/A = 50 MeV}

Before discussing the influence of the symmetry potential and the in-medium NN cross sections on various HIC observables, we first show some predictions for dynamical processes and fragment production mechanisms in $\mathrm{Sn+Sn}$ collisions at $\mathrm{E/A=50MeV}$. Figs. 2 (a) and (b) show the time evolution of the density contour plots for $\mathrm{^{124}Sn+^{124}Sn}$ at $E/A=50MeV$ for $b=0$ (top panels) and $b=6 fm$ (bottom panels) from one typical event. Here, $\gamma_i=0.5$ was used in the calculations, but the density contour plots are practically the same for all $\gamma_i>0.2$. (For smaller $\gamma_i$, many more nucleons are emitted during simulations due to the instability of the initial nuclei.)
The projectile and target touch around $50fm/c$ and nucleons start to transfer between the two nuclei. The compressed region reaches the highest density at approximately $100 fm/c$ for both impact parameters. However, higher densities are reached in the overlap region and more compressional energy is stored for calculations with the soft-asy EoS potentials than with the stiff-asy EoS potentials.

In central collisions ($b=0fm$) at an incident energy of $E/A=50MeV$, the system reaches a maximum density at about 100fm/c and subsequently expands to low density where multifragmentation occurs. At this point, the reaction system with the soft-asy EoS disintegrates into more light fragments than the system with the stiff-asy EoS. In peripheral collisions ($b=6 fm$), the projectile and target residues separate after $150fm/c$ has elapsed, whereupon the low-density neck connecting them ruptures into fragments. The two excited residues then continue along their paths without further mutual interactions. Thus the nucleon diffusion process in peripheral heavy ion collision is predicted to terminate at about $150fm/c$.
 \begin{figure*}[htbp]
     \centering
 \includegraphics[angle=270,width=0.8\textwidth]{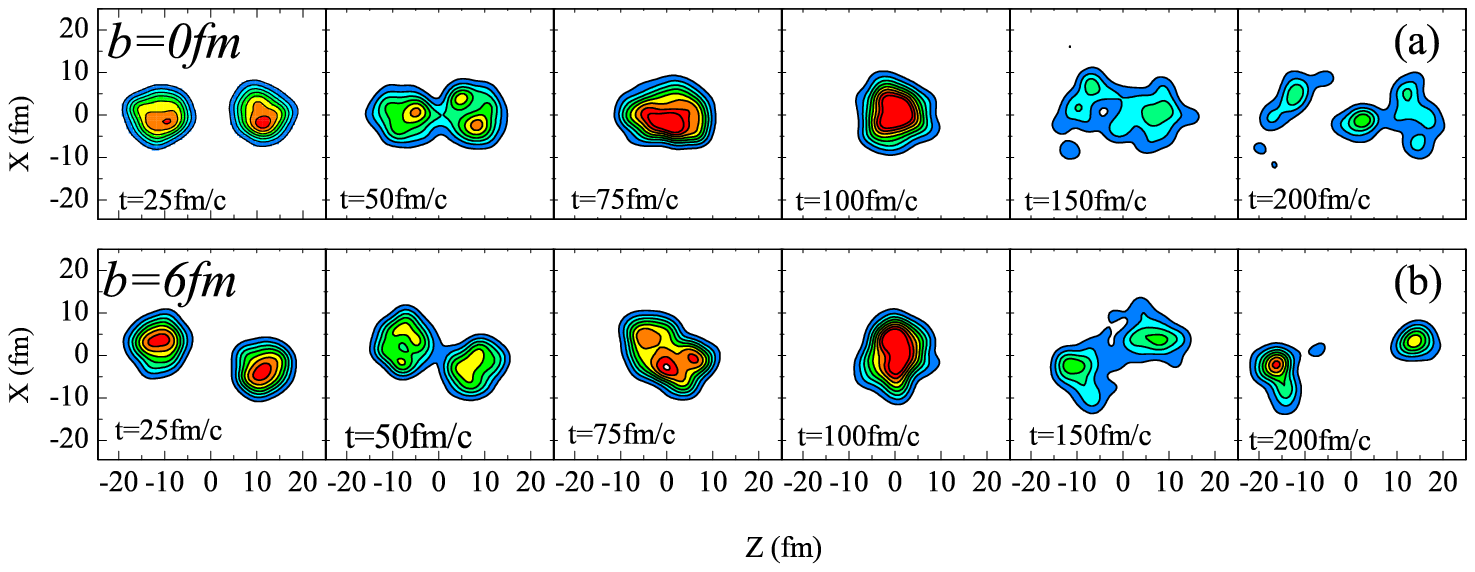}%
 \setlength{\abovecaptionskip}{20pt}
 \caption{(Color online) Time evolution of the nucleon density in the reaction plane for$\mathrm{^{124}Sn+^{124}Sn}$ collisions at E/A=50MeV at b=0 (top panels) and 6 (bottom panels) fm.}
  \setlength{\belowcaptionskip}{40pt}
 \end{figure*}

Fig. 3 shows the charge (left panels) and mass (right panels) distributions of the reaction products for $\mathrm{^{112}Sn+^{112}Sn}$ at an incident energy of $50MeV$
per nucleon for $b=2, 6, 8 fm$. The solid symbols are the results for $\gamma_i=2.0$ and open circles are the results for $\gamma_i=0.5$. The charge and mass distributions depend on the symmetry potential parameter $\gamma_i$. The distributions obtained with $\gamma_i=0.5$ are narrower than those obtained with $\gamma_i=2.0$.

For completeness, the multiplicities of fragments with charge Z as a function of their scaled rapidity $y/y^{c.m.}_{beam}$ for $\mathrm{^{112}Sn+^{112}Sn}$ and $b=2, 6, 8 fm$ are shown in Fig. 4. The maximum
charge of fragments at mid-rapidity decreases with increasing impact parameter $b$. In general, two ridges of heavier fragments are observed, distributed near the initial projectile and target rapidities. For reference, the initial projectile and target rapidities are marked by the vertical dotted lines at $y/y^{c.m.}_{beam}=1$ and $-1$, respectively. The heaviest fragments have lost about $35\%$ of their initial
velocity for central collisions, and about $25\% (10\%)$ of their initial velocity for $b=6fm$ ($b=8fm$). The velocity loss of the heaviest fragments depends on the decelerating effects from the
effective N-N interactions and nucleon-nucleon collision frequency. These features have been observed in experiments \cite{TXLiu07,Nebau99}.  Intermediate mass fragments with Z between 3 and 20 are produced over a wide range of rapidities for every impact parameter.

 \begin{figure}[htbp]
 \centering
 \includegraphics[angle=270,width=0.4\textwidth]{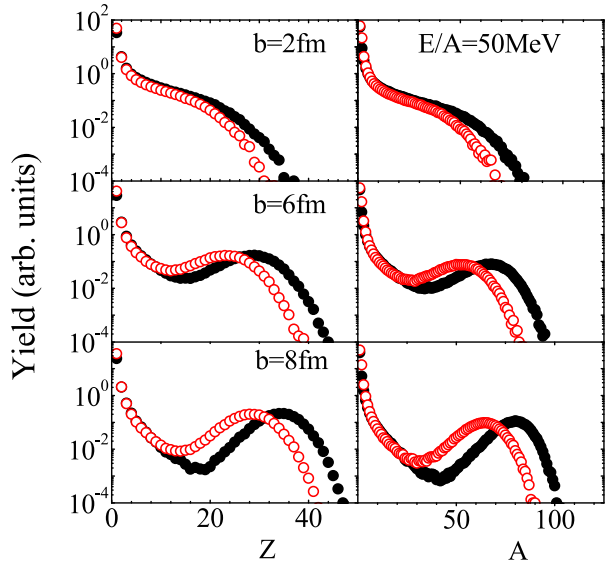}%
 \setlength{\abovecaptionskip}{40pt}
 \caption{(Color online) The charge (left panels) and mass (right panels) distributions for $\mathrm{^{112}Sn+^{112}Sn}$ at b=2 (top panels), 6 (middle panels), and 8 (bottom panels) fm. The solid symbols are for $\gamma_i=2.0$ and open symbols are for $\gamma_i=0.5$.}
  \setlength{\belowcaptionskip}{40pt}
 \end{figure}

 \begin{figure}[htbp]
     \centering
 \includegraphics[angle=270,width=0.5\textwidth]{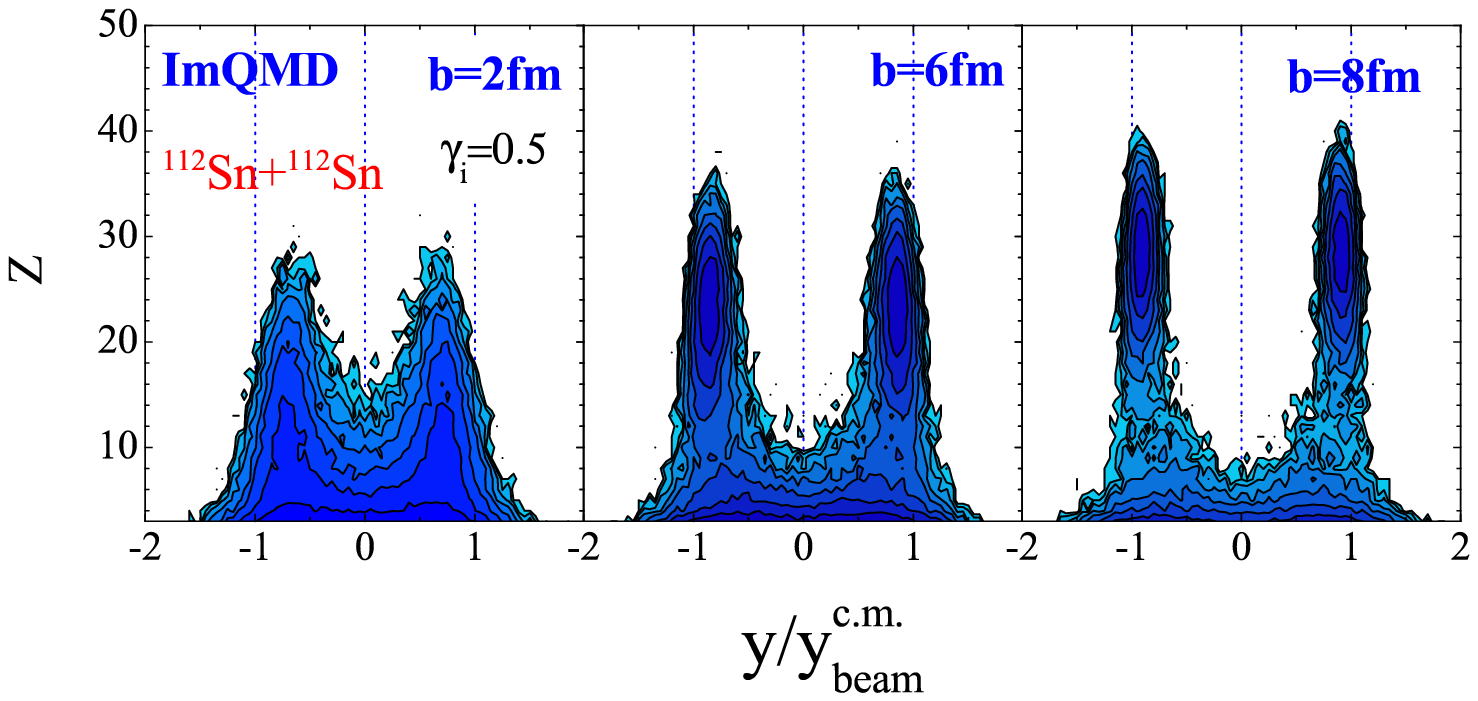}%
 \setlength{\abovecaptionskip}{20pt}
 \caption{(Color online) Multiplicity of fragments with charge Z as a function of their scaled rapidity $y/y^{c.m.}_{beam}$ for $\mathrm{^{112}Sn+^{112}Sn}$ at E/A=50MeV for b=2 fm (left panel), 6 fm (middle panel), and 8 fm (right panel).}
  \setlength{\belowcaptionskip}{40pt}
 \end{figure}

\section{Isospin observables}
In this section, we examine the uncertainties in constraining the density dependence of the symmetry energy from transport model simulations.  To do this, we study the effect of varying the isospin dependence of in-medium NN cross-sections, the impact parameter, and the symmetry potential on isospin sensitive observables. In particular, we focus on the calculated results of the neutron to proton yield ratio as well as the isospin transport ratio ($R_{i}$). These two observables have been used extensively to provide constraints on the density dependence of symmetry energy at subsaturation density\cite{BALi97, BALi05, LWChen05,Rizzo08}.

\subsection{Nucleon yield ratios}
The symmetry potential, which is opposite in sign for protons and neutrons, directly affects the emission of nucleons. This fact
was recognized early on and has been studied quite intensely \cite{BALi97,BALi06,Tsang09,zhang08,Pfabe08}.
Experimentally, due to their low interaction with matter, neutrons are more difficult to measure. Furthermore, different types of
detectors are usually employed to measure neutrons and protons. To minimize the effects of detector efficiencies, the yield ratio from an asymmetric system A is compared to the yields from a more symmetric system B by constructing
the double ratio $DR(n/p)=R_{n/p}(A)/ R_{n/p}(B)=(Y_{n}(A)/Y_{p}(A))/(Y_{n}(B)/Y_{p}(B))$,
where $Y_{n}(A)$ and $Y_{p}(A)$ are the neutron and proton yields obtained in system $A$, and $Y_n(B)$ and $Y_p(B)$ are the neutron and proton yields obtained in system B. In Fig. 5, the $\mathrm{DR(n/p)}$
data from two collision systems $\mathrm{A=^{124}Sn+^{124}Sn}$ and $\mathrm{B=^{112}Sn+^{112}Sn}$ for central collisions are
plotted as solid stars in the left panel. The shaded regions in Fig. 5 are the ImQMD05 results
for two different symmetry energy cases $\gamma_i=0.5$ (upper shaded region) and $2.0$ (lower shaded region), where the
transverse emitted nucleons have been selected with an angular cut of $70^{o}<\theta_{c.m.}<110^{o}$ as in Ref. \cite{Fami06,zhang08}.

The $\mathrm{DR(n/p)}$ values for
the $\gamma_i=0.5$ case are greater than that for $\gamma_i=2.0$. In intermediate energy HICs, the emitted nucleons mainly come from the overlap region during the expansion phase.  This region is below saturation density, where the symmetry energy is larger for smaller $\gamma_i$. Larger symmetry
energy in this region results in enhanced neutron emissions from the neutron-rich system and thus larger values of the double ratio. This
behavior has been observed in other transport model predictions\cite{BALi97,BALi06,Pfabe08}. However, some of these other calculations using BUU approach produce double ratios that are much smaller than the data for all symmetry energies, while the ImQMD05 calculations for $\gamma_i=0.5$ are in the correct range. It remains a puzzling discrepancy between transport model predictions, and detailed comparisons between the different codes and different transport models should be done in the future to understand this difference better. Such study would allow more definite constraints on the symmetry energy from transport models.

The impact parameter dependence of $DR(n/p)$ ratios is plotted in the right panel of Fig. 5. The energy cut, $40 \le E_{c.m.} \le 80MeV$, and angular gates are chosen to minimize sequential decays and cluster effects on the calculated values. The cuts also provide more robust coalescence-invariant quantities. The impact parameter dependence is rather weak. (Noting that there is a large zero-offset on the y-axis.) One reason for the insensitivity with respect to impact parameter in these calculations is that transverse emitted nucleons mainly come from the overlap region, where the high energy values of $\mathrm{DR(n/p)}$ are related to the information of symmetry energy of nuclear matter, regardless of the size of that region.

 \begin{figure}[htbp]
     \centering
 \includegraphics[angle=90,width=0.45\textwidth]{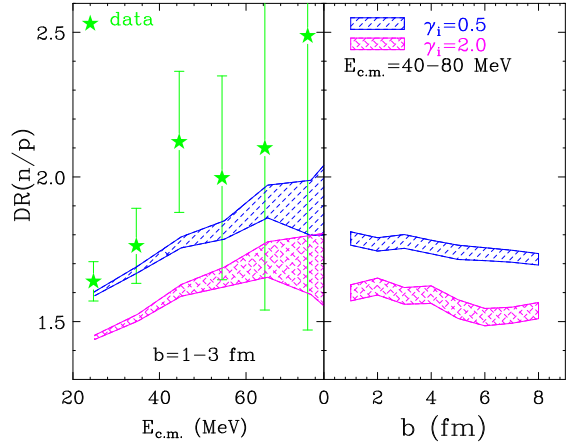}%
 \setlength{\abovecaptionskip}{20pt}
 \caption{(Color online) (Left panel) Double n/p yield ratios, DR(n/p), from transverse emitted nucleons as a function of kinetic energy. The solid stars are data from Ref. \cite{Fami06}. (Right panel) DR(n/p) from transverse emitted high kinetic energy nucleons ($E_{c.m.}>40MeV$) as a function of impact parameter. The upper shaded region is for $\gamma_i=0.5$ and the lower shaded region is for $\gamma_i=2.0$.}
  \setlength{\belowcaptionskip}{0pt}
 \end{figure}

Similar to the symmetry potential, the isospin dependence of in-medium nucleon-nucleon (NN) cross sections should play an important role on the isospin sensitive observables for heavy ion collisions in the transport model. The in-medium NN cross sections depend on the type of colliding particle, the relative energy of the colliding pair and the local density of medium. The calculations are performed at b=2fm where the sensitivity to the in-medium cross section are relatively larger because the density is higher on the average and there are more nucleon-nucleon collisions for central nucleus-nucleus collisions. We investigate three options for the in-medium NN cross sections :1) we used the free nucleon-nucleon cross sections, with no reduction $\sigma^{*}_{nn/np}=\sigma^{free}_{nn/np}$; 2) We used an experimentally motivated $\sigma^{*}_{NN}$ reduction, $\sigma^{*}_{nn/np}=(1-0.2\rho/\rho_{0})\sigma^{free}_{nn/np}$;  3) We used isospin independent reduced $\sigma^*_{nn/np}$ case, $\sigma^{*}_{nn}=\sigma^{*}_{pp}=\sigma^{*}_{np}=(1-0.2\rho/\rho_{0})\sigma'$, where $\sigma'=(2N_{np}\sigma^{free}_{np}+(N_{nn}+N_{pp})\sigma^{free}_{nn/pp})/N_{NN}$.  In case 3, $N_{np}$, $N_{nn/pp}$, $N_{NN}$ are the number of possible combinations of np, nn, pp, and NN colliding pairs in the reaction system.  By keeping the total number of NN collisions to be almost the same when we compare calculations with case 2 to those with case 3, we can study the effect of the isospin dependence of the in-medium cross sections without varying the overall nucleon-nucleon collision rate. Here, we are explicitly interested in the isospin dependence.

The calculated results of the $\mathrm{DR(n/p)}$ ratios are plotted in Fig.6 for the energy and angular gates described above, which emphasize the emission from the overlap region formed during the early stages of the HICs. By comparing case 2 (open squares) to case 3 (solid circles), we can test whether it makes any difference to the double ratios, whether the overall collision rate is dominated by n-p collisions as one would expect if the in-medium reduction is the same for all cross sections, or whether n-n, p-p and n-p collisions are governed by the same in-medium cross sections. The effect of reducing the cross sections below their free values can be investigated by comparing the DR(n/p) results obtained with case 1 (inverted triangles) and case 2 (open circles). The changes in the cross section have no effect for $\gamma_i$=0.5. For $\gamma_i=2.0$, there is a slight difference between case 1 and case 2, indicating some influence of the total number of collisions when the symmetry potential is smaller at subsaturation density.  However, the effect of the isospin dependence of the in-medium NN cross sections on the $DR(n/p)$ is smaller than the statistical errors for both $\gamma_i$ we studied. Overall, the $\mathrm{DR(n/p)}$ exhibits little to no dependence on the in-medium NN cross section.  The main reason is that NN collisions are suppressed by Pauli blocking at $E/A=50MeV$. The different $\sigma^{*}_{nn/np}/\sigma^{free}_{nn/np}$ or $\sigma^{*}_{np}/\sigma^{*}_{nn}$ values used in the transport model do not lead to significant differences in the $\mathrm{DR(n/p)}$ ratios. Thus, the isospin observable $\mathrm{DR(n/p)}$  from Sn+Sn at $E/A=50MeV$ is a robust observable to constrain the density dependence of symmetry energy.
 \begin{figure}[htbp]
     \centering
 \includegraphics[angle=90,width=0.45\textwidth]{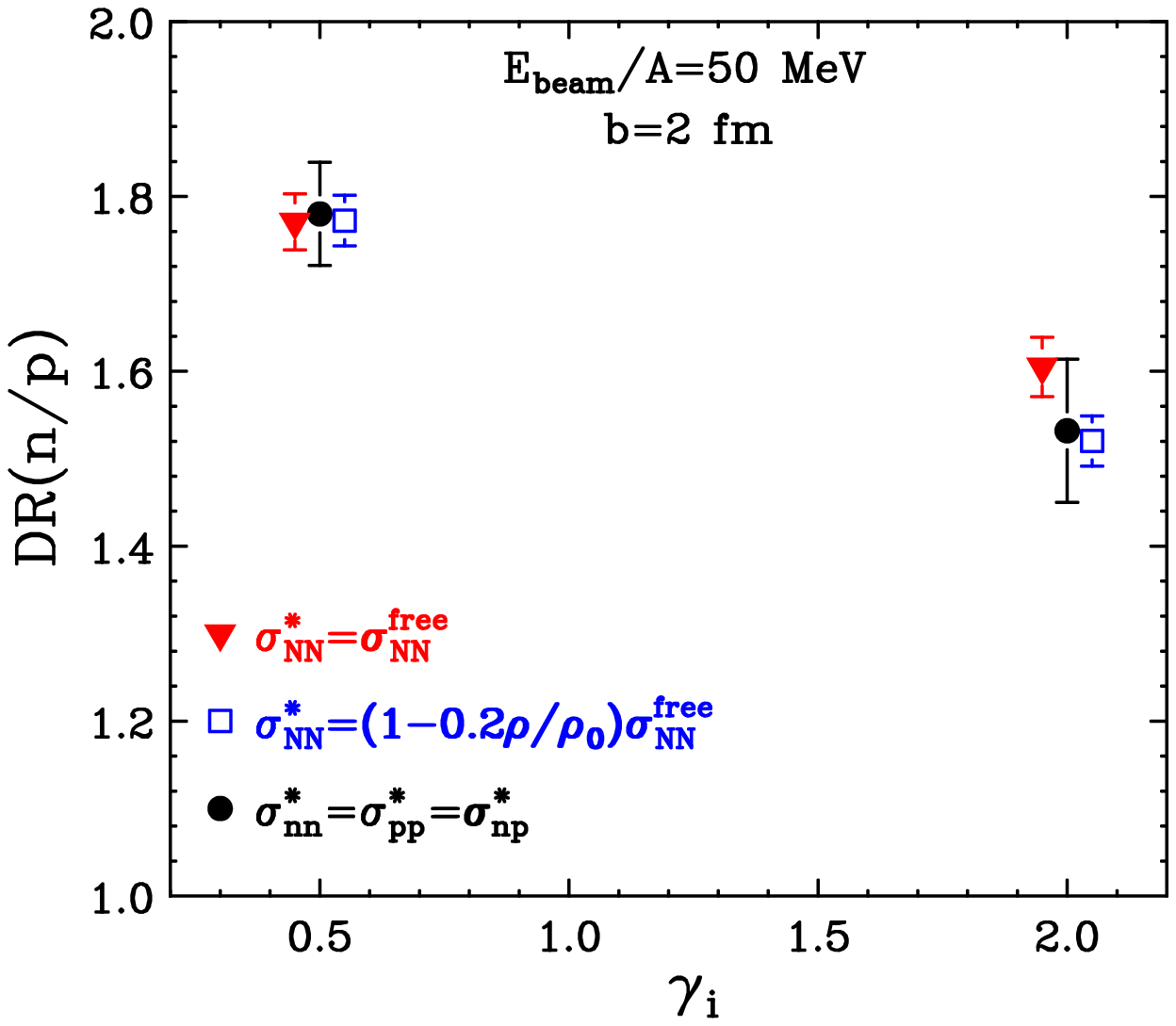}%
 \setlength{\abovecaptionskip}{20pt}
 \caption{(Color online) Effects of different in-medium NN cross sections on the DR(n/p) from transverse emitted high energy nucleons ($E_{c.m.}>40MeV$). The inverted triangles are for the free space NN cross section (case 1 in text). The open squares are for the isospin-dependent in-medium NN cross section with a phenomenological formula   (case 2). The solid circles are for the isospin-independent in-medium NN cross section (case 3). }
  \setlength{\belowcaptionskip}{0pt}
 \end{figure}

\subsection{Isospin transport ratio $R_i$}

When the projectile and target nuclei come into contact, there can be exchange of nucleons between them. If the neutron to proton ratios of the projectile and target differ greatly, the net nucleon flux can cause a diffusion of the asymmetry $\delta$ reducing the difference between the asymmetries of the two nuclei. This isospin diffusion process, which depends on the magnitude of the symmetry energy, affects the isospin asymmetry of the projectile and target residues in peripheral HICs. The isospin transport ratio $R_i$ has been introduced \cite{Tsang04} to quantify the isospin diffusion effects,
\begin{equation}
R_i=\frac{2X-X_{aa}-X_{bb}}{X_{aa}-X_{bb}}, \label{Ridef}
\end{equation}
where X is an isospin observable and the subscripts $a$ and $b$ represent the neutron rich and neutron-poor nuclei.  In this work, we use $a$ and $b$ to denote the projectile (first index) and target (second index) combination. where $\mathrm{a=^{124}Sn}$, and $\mathrm{b=^{112}Sn}$. We obtain the value of $R_{i}$ by comparing three reaction systems, $\mathrm{a+a}$, $\mathrm{b+b}$ and $\mathrm{a+b}$ (or $\mathrm{b+a}$). Construction of the transport ratio minimize the influence of other effects besides isospin diffusion effects on the fragment yields, such as preequilibrium emission and secondary decay, by rescaling the observable X for the asymmetric a+b system by its values for the neutron-rich and neutron-deficient symmetric systems, which do not experience isospin diffusion. Based on Eq. (\ref{Ridef}), one expects $R_i=\pm1$ in the absence of isospin diffusion and $R_i\sim0$ if isospin equilibrium is achieved. Eq. (\ref{Ridef}) also dictates that two different observables, $\mathrm{X}$, will give the same results if they are linearly related.
In one experiment, $\mathrm{X}$ was taken as the isoscaling parameter, $\alpha$, obtained from the yield of the light particles near the projectile rapidity\cite{Tsang01}. In transport models \cite{Tsang04,LWCh04} based on the Boltzmann Uehling Uhlenbeck (BUU) approach, the isospin asymmetry $\delta$ of the projectile residues (emitting source) has been used to compute $R_i(\delta)$. The constraints on the density dependence of symmetry energy were obtained by comparing BUU predictions to the isospin diffusion data\cite{Tsang04,LWCh04,BALi05,Rizzo08}. This is possible only if the isoscaling parameters $\alpha$ are linearly related to the isospin asymmetry, $\delta$, of the projectile residues. This has been shown both experimentally and theoretically \cite{TXLiu07,Tsang01,Ono03} in central collisions.

In the BUU analysis, the projectile residues are defined as the sum of regions with local density greater than $0.05\rho_{0}$ and center of mass velocity greater than half of the beam velocity \cite{Tsang04}.  The definition of a residue based on this criterion is not applicable to the QMD model as a clustering process is introduced and these residues disintegrate, producing fragments. In our ImQMD05 calculations at $b=6fm$ and $E/A=50MeV$, we find that isospin diffusion for $\mathrm{^{124}Sn+^{112}Sn}$ at $b=6fm$ stops around $150fm/c$. The emission of nucleons and fragments at later stages is not part of the isospin diffusion process, but may change the isospin contents of the final fragments and thus needs to be included in the analysis in order to accurately quantify the diffusion process. In the following, we calculate the collisions of four reaction system, $\mathrm{^{124}Sn+^{124}Sn}$, $\mathrm{^{124}Sn+^{112}Sn}$, $\mathrm{^{112}Sn+^{124}Sn}$, $\mathrm{^{112}Sn+^{112}Sn}$ at incident energy of $50 MeV$ per nucleon for variety of parameters. We use two different isospin tracers to calculate the isospin transport ratios of Eq. (\ref{Ridef}). Our studies lead us to propose another isospin transport ratio for constraining the symmetry energy in future experiments.

We first analyze the amount of isospin diffusion by constructing a tracer from the isospin asymmetry of all emitted nucleons (N) and fragments (frag), including the heavy residue if it exists, with velocity cut $v^{N,frag}_z>0.5v^{c.m.}_{beam}$ (nearly identical results are obtained with higher velocity cut $v^{N,frag}_z> 0.7v^{c.m.}_{beam}$).  This represents the full projectile-like emitting source, and should be comparable to what has been measured in experiments.  Fig. 7 shows the results of isospin transport ratios $R_i(X=\delta_{N,frag})$ (upright triangles) as a function of the impact parameter for a soft symmetry case ($\gamma_i=0.5$, open symbols) and a stiff symmetry case ($\gamma_i=2.0$, closed symbols). $R_i$ obtained with soft-symmetry case is smaller than those obtained with stiff-symmetry potential case. This is consistent with the expectation that higher symmetry energy at subnormal density leads to larger isospin diffusion effects (smaller $R_i$ values).
 \begin{figure}[htbp]
     \centering
 \includegraphics[angle=90,width=0.40\textwidth]{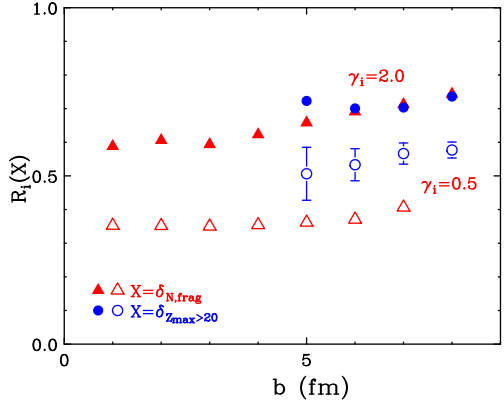}%
 \setlength{\abovecaptionskip}{20pt}
 \caption{(Color online) Isospin transport ratios as a function of impact parameter with two tracers for a soft symmetry case ($\gamma_i=0.5$, open symbols) and a stiff symmetry case ($\gamma_i=2.0$, closed symbols). Upright triangle symbols are for the tracer defined by the isospin asymmetry of all fragments and unbound nucleons with velocity cut ($v^{N,frag}_z>0.5v^{c.m.}_{beam}$), $X=\delta_{N,frag}$. Circles are for the tracer defined by the heaviest fragment with $Z_{max} > 20$ in projectile region, $X=\delta_{Z_{max} > 20}$.}
  \setlength{\belowcaptionskip}{0pt}
 \end{figure}

$R_i$ depends weakly on impact parameter over a range extending from central ($b=3fm$) to mid peripheral($b=8fm$) collisions. Interestingly, the isospin equilibrium and global thermal equilibrium are not reached even for central collisions. It is different than the results obtained in \cite{Rizzo08}. Our results show, that neither the effective interaction is sufficiently strong nor the collisions are sufficiently frequent (most of them are Pauli suppressed) to mix the projectile and target nucleons completely. These two effects prevent the combined system from attaining isospin equilibrium even in central collisions. With impact parameter increasing for $b>5fm$, the overlap region and thus the number of nucleons transferred from projectile and target decreases, causing the $R_i$ values to increase. The impact parameter dependence of $R_i$ predicted by ImQMD05 has been compared to experimental data at 35AMeV where the measured trends are more consistent with the softer symmetry energy ($\gamma_i=0.5$) than with the stiffer symmetry energy ($\gamma_i=2.0$)\cite{sun10}.

In peripheral collisions, most often, a large residue remains. If it decouples from the full emitting source before it equilibrates, it may experience a different amount of diffusion than the full emitting source examined by $X=\delta_{N,frag}$. To examine this, we constructed a tracer using the isospin asymmetry of the heaviest fragments with charge $Z_{max}>20$ in the projectile region. This tracer is mainly relevant to peripheral collisions as the central collisions are dominated by multifragmentation and very few large projectile fragments survive. The dependence of $R_i(X=\delta_{Z_{max}>20})$ for impact parameter $b\geq5fm$ is shown as open and closed circles in Fig. 7. The isospin transport ratios constructed from the different isospin tracers have different values especially in the case of $\gamma_i=0.5$. Stronger isospin equilibration (smaller $R_i$ values) is observed in the isospin transport ratios $R_i(X=\delta_{N,frag})$ constructed from nucleons and fragments than $R_i(X=\delta_{Z_{max}>20})$ constructed from the heaviest fragments with $Z_{max} > 20$. Since isospin diffusion mainly occurs through the low-density neck region,
and the system breaks up before isospin equilibrium, the asymmetry of the projectile and target residues do not achieve equilibrium and, larger $R_i(X=\delta_{Z_{max}\ge20})$ values result. In contrast, there is more mixing of nucleons from the target and projectile in the neck region due to the isospin diffusion. Consequently, rupture of the neutron-rich neck is predicted to result in the production of neutron-rich fragments at mid rapidity. Experiments \cite{expD} have been proposed in both Rare Isotope Beam Factory at RIKEN and the Coupled Cyclotron Facility at NSCL to test these predictions by comparing the isospin transport ratios obtained from residues and from intermediate mass fragments at different rapidity values.

Since fragments are formed at all rapidities, we can examine the rapidity dependence of $R_i$ to obtain more information about the reaction dynamics. Fig. 8 shows $R_i$ as a function of the scaled rapidity $y/y_{beam}$. The symbols in the leftmost panel are experimental data obtained in Ref. \cite{TXLiu07} for three centrality gates. This transport ratio was generated using the isospin tracer $X=ln(Y(^7Li)/Y(^7Be))$ where $Y(^7Li)/Y(^7Be)$ is  the yield ratio of the mirror nuclei, $^7Li$ and $^7Be$ \cite{TXLiu07}. As expected the values of $R_i$ obtained from peripheral collisions (solid stars) are larger than those obtained in central collisions (open stars). For comparison, the ImQMD05 calculations of $R_i(X=\delta_{N,frag})$ are plotted as lines in the middle and right panels for a range of impact parameters. The middle panel contains the results from the soft symmetry potential ($\gamma_i=0.5$) while the right panel shows the results from the stiff symmetry potential ($\gamma_i=2.0$). The impact parameter trends and magnitude of the data are more similar to the results of the calculations from soft symmetry potentials ($\gamma_i=0.5$), consistent with previous analysis \cite{Tsang09}. However, the experimental trend of $R_i$ gated on the most central collisions (open stars) is not reproduced by the calculations. The experimental data indicate more equilibration for central collisions near mid rapidity while the transport model indicates more transparency. The equilibration in the E/A = 50MeV data may be the result of the impact parameter determination from charged particle multiplicity wherein the most central collisions are assumed to be the ones with highest charge particle multiplicity. For the most central events, a gate on the highest multiplicity, may select events in which more nucleon-nucleon collisions occur rather than a strict selection on the most central impact parameters. We plan to study the impact parameter smearing effect in detail in a future paper.

 \begin{figure}[htbp]
     \centering
 \includegraphics[angle=90,width=0.45\textwidth] {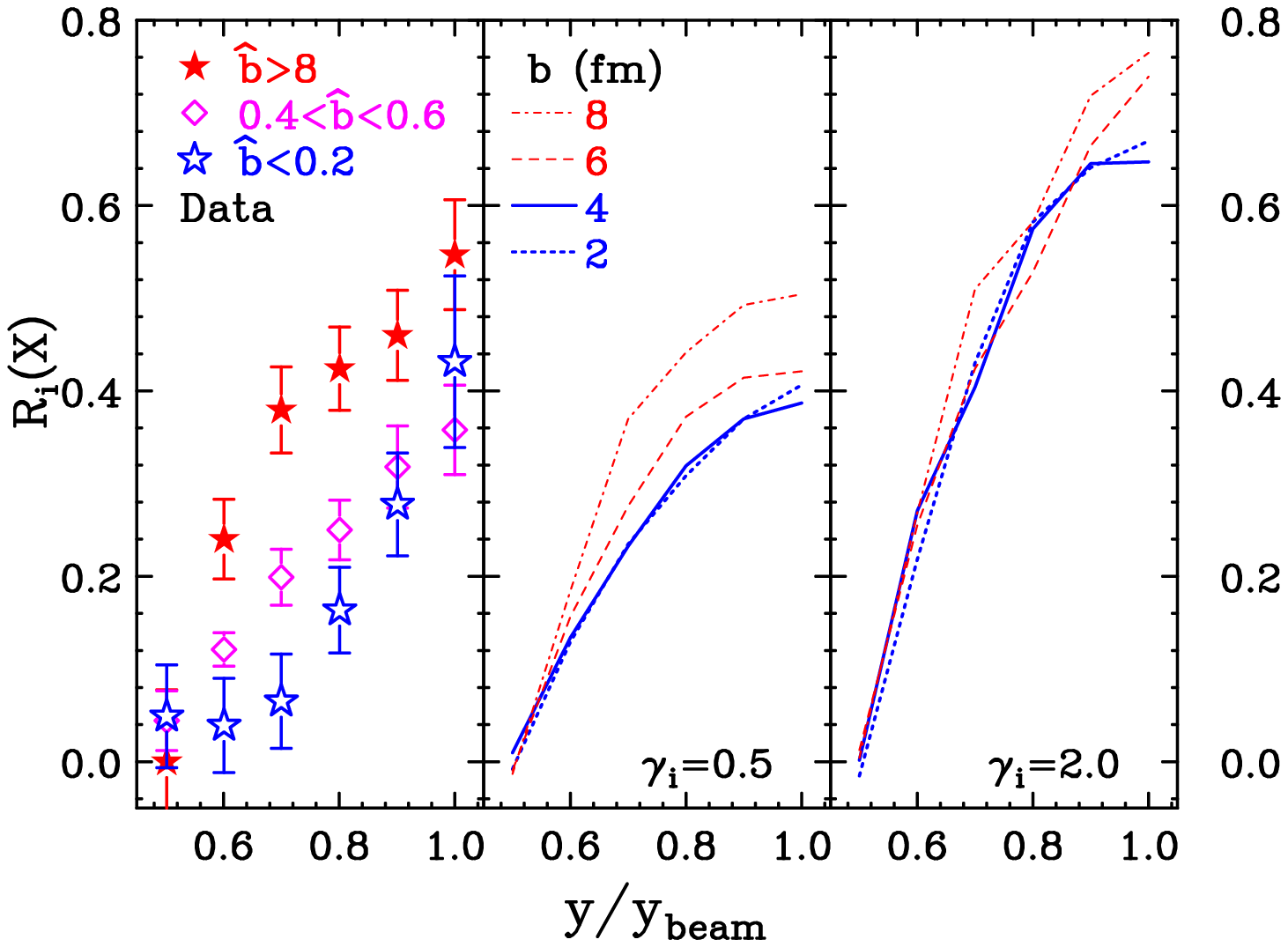} 
 \setlength{\abovecaptionskip}{20pt}
 \caption{(Color online) (Left panel) Experimental $R_i$ as a function of rapidity for three centrality gates [16]. (Middle panel) The calculated results of $R_i(X=\delta_{N,frag})$ as a function of rapidity for $b= 2, 4, 6, 8 fm$ for $\gamma_i=0.5$ and (Right panel) $\gamma_i=2.0$.}
  \setlength{\belowcaptionskip}{0pt}
 \end{figure}

Next, we discuss the influence of the isospin dependence of the in-medium nucleon-nucleon cross section on isospin diffusion, using the three cases described in Section III.A. Previous studies have shown that larger in-medium NN cross sections enhance isospin diffusion \cite{Shi03}, and so the value of $R_i$ in case 1 (free cross sections, inverted triangles) should be smaller than the $R_i$ values for case 2 (reduced cross sections, open squares). But as Fig. 9 shows, there is no noticeable effect at the energy we studied. Two effects lead to this result. First, the dynamical process of heavy ion collisions is governed more by the mean field than the nucleon-nucleon collisions at $50 MeV$ per nucleon. Second, the nucleon-nucleon collision frequency is small for peripheral HICs. The direct comparison between the results from BUU and ImQMD is not appropriate as the two calculations use very different density dependence of symmetry energy and in-medium nucleon-nucleon cross section. In fact when similar range of symmetry stiffness is used in both IBUU04 (x=-1 and x=0) and ImQMD05 ($\gamma_i$=0.5 and 2.0) as shown in Figure 1, the magnitude of the $R_i$ values are similar. Any remaining differences could be caused by the differences on the exact form of in-medium NN cross sections and Pauli blocking adopted in the simulations. Indeed, both transport models predict that isospin diffusion depends more strongly on the symmetry potential and less strongly on the isospin dependence of the in-medium NN cross section at this beam energy.
\begin{figure}[htbp]
     \centering
 \includegraphics[angle=90,width=0.45\textwidth] {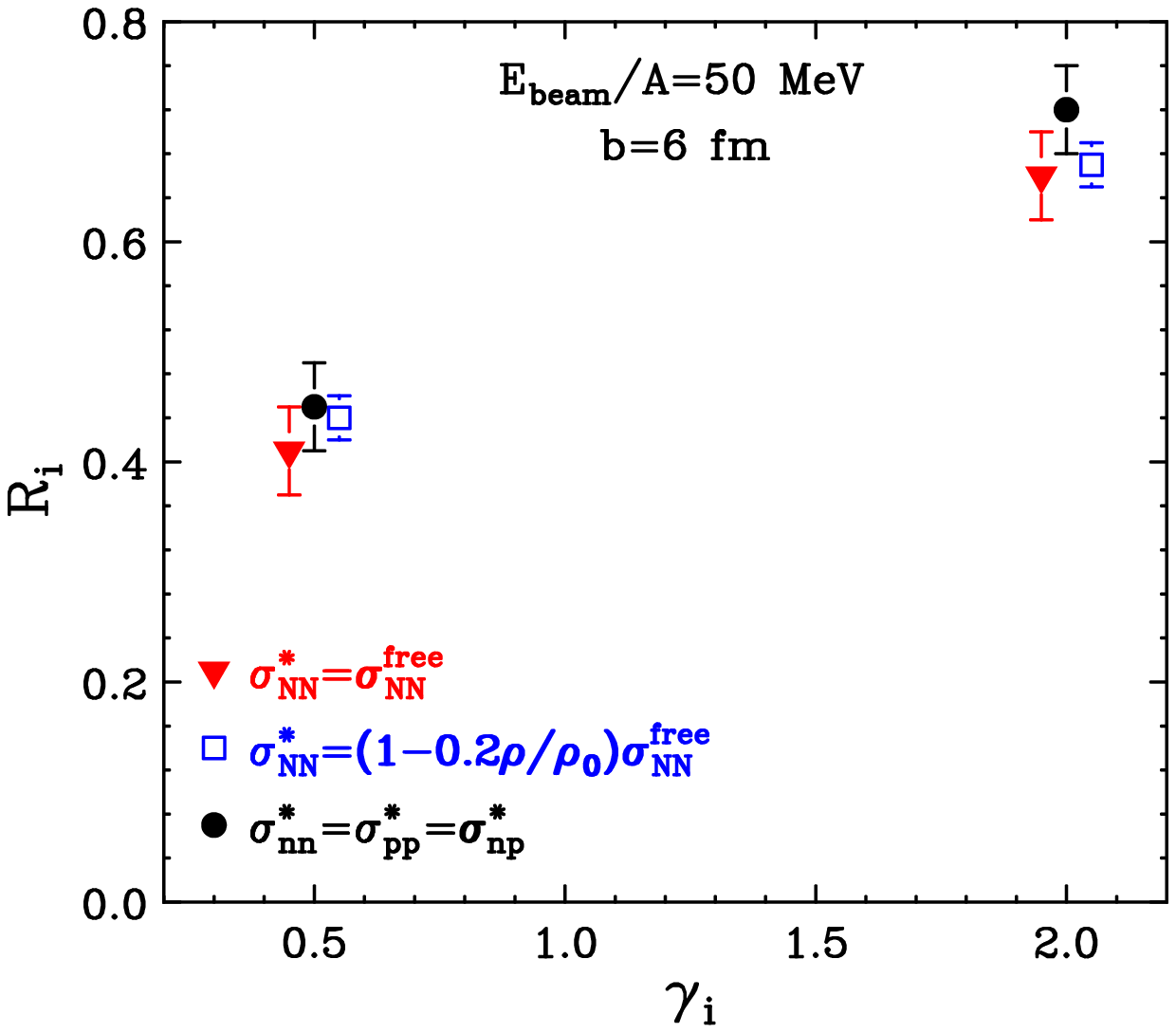} 
 \setlength{\abovecaptionskip}{20pt}
 \caption{(Color online) The effects of different in-medium NN cross sections on $R_i(X=\delta_{N,frag})$. The inverted triangles are for the free space NN cross section  $\sigma^{*}_{nn/np}=\sigma^{free}_{nn/np}$ (case 1 in text).  The open squares are for the isospin-dependent in-medium NN cross section with phenomenological formula $\sigma^{*}_{nn/np}=(1-0.2\rho/\rho_{0})\sigma^{free}_{nn/np}$ (case 2). The solid circles are for the isospin-independent in-medium NN cross section $\sigma^{*}_{nn}=\sigma^{*}_{pp}=\sigma^{*}_{np}$ (case 3).}
  \setlength{\belowcaptionskip}{0pt}
 \end{figure}

\section{summary}
In summary, we have investigated the influences of the density dependence of the symmetry energy, of the in-medium nucleon-nucleon cross section and of the impact parameters on several different isospin observables predicted by ImQMD05. The study shows that these isospin dependent observables are more strongly influenced by the mean field than by the NN collisions. In particular, the double n/p yield ratios ($\mathrm{DR(n/p)}$) and the isospin transport ratios ($R_i$) are more sensitive to the density dependence of the symmetry energy than to the isospin dependence of the in-medium NN cross sections. This conclusion is similar to conclusions reached using BUU approaches in the range of symmetry energies studied here.  The fact that two different approaches used to simulate isospin diffusion draw the same conclusions indicates real progress on constraining the symmetry energy with heavy ion collisions.

We also examine the impact parameter dependences of the $\mathrm{DR(n/p)}$ ratio and of the isospin transport ratio $R_i$. Our results illustrate that there is only a weak impact parameter dependence of $\mathrm{DR(n/p)}$ from central to peripheral collisions.  Likewise, there is only a weak impact parameter dependence of $R_i$ from central to mid-peripheral collisions. From mid-peripheral to peripheral collisions, however, the predicted $R_i$ values increase rapidly with impact parameter reflecting the decreasing number of nucleons transferred between projectile and target in increasingly peripheral collisions.

Cluster formation is important for intermediate energy heavy ion collisions. The ImQMD05 approach produces clusters at all rapidities, which allows us to study reaction dynamics by means of $R_i$ as a function of rapidity. We find that the ImQMD05 approach describes the data better with soft symmetry energies term than with stiff symmetry energies. We also tested different tracers by constructing corresponding isospin transport ratios for them using different symmetry energies. For weakly density dependent symmetry energies (small $\gamma_i$) with large symmetry energies at sub-saturation densities, the values of $R_i$ for the residue tracer $X=\delta_{Z_{max}>20}$ are larger than those extracted from the entire emitting source, i.e., $X=\delta_{N,frag}$. Calculations using the latter tracer has been previously compared to the measured isospin diffusion data in heavy ion collisions. The difference between these two tracers can be examined experimentally as a new probe of the symmetry energy.

\begin{acknowledgements}
This work has been supported by the Chinese National Science Foundation under Grants 11075215, 10979023,10675172, 11005022, 10235030 and the U.S. National Science Foundation under Grants PHY-0216783, 060007, 0800026, the High Performance Computing Center (HPCC) at Michigan State University and the Texas Advanced Computing Center, University of Texas.
\end{acknowledgements}

\appendix*

\section{I}
The explicit momentum dependent interaction (MDI) term introduced in Eq. (\ref{eq:umd}) contributes to the EoS at T=0MeV as follows:
\begin{eqnarray}
E_{md}/A&&=\varepsilon_{md}/\rho=\nonumber \\
&&\frac{1.57}{36\epsilon^3}u[18(u^2\epsilon^3+1)\log^2(u^{2/3}\epsilon+1)\nonumber \\
&&-6(2u^{2}\epsilon^3-3u^{4/3}\epsilon^2+6u^{2/3}\epsilon+11)\log(u^{2/3}\epsilon+1))\nonumber \\
&&+(4u^2\epsilon^3-15u^{4/3}\epsilon^2+66u^{2/3}\epsilon)]
\end{eqnarray}
Here $u=\rho/\rho_0$, $\epsilon=21.57$ and the energy is in $MeV$. This form of the momentum dependent interaction changes the mean field potential and therefore the EoS. Eq. (\ref{eq:umd}) should then have a correction to recover the EoS obtained with the Skyrme interaction parameters at T=0MeV, which is not included in our calculations in this paper.  To see the influence of this correction, we can add a counter term  as follows.
\begin{eqnarray}
u_{md}=\frac{1}{2\rho_{0}}\sum_{N_1,N_2} && \frac{1}{16\pi^{6}}\int d^{3}p_{1}d^{3}p_{2}f_{N_{1}}(\vec{p}_1)f_{N_{2}}(\vec{p}_2)\nonumber\\
&&1.57[\ln(1+5\times 10^{-4}(\Delta p)^2)]^2 \nonumber\\
-\frac{1}{2\rho_{0}}\sum_{N_1,N_2} && \frac{1}{16\pi^{6}}\int^{p_f} d^{3}p_{1}d^{3}p_{2}f^0_{N_{1}}(\vec{p}_1)f^0_{N_{2}}(\vec{p}_2)\nonumber\\
&& 1.57[\ln(1+5\times 10^{-4}(\Delta p)^2)]^2 \label{umd1}
\end{eqnarray}
Here, $f_N(\vec{p}_1)$ are the nucleon Wigner functions at a given temperature, and the $f^0_N(\vec{p}_1)$ represent the nucleon Wigner density at T=0MeV.  The second term is just the first term evaluated at zero temperature, so with this correction to the momentum dependent interaction we recover the EoS of cold nuclear matter (T=0MeV) obtained with the Skyrme interaction parameters.

When we repeat the simulations of the $\mathrm{Sn+Sn}$ collisions using the momentum dependent interaction of Eq. (\ref{umd1}), albeit with less statistics, the results of the mass/charge distributions, neutron/proton ratios and isospin transport ratios are the same as the results obtained using Eq. (\ref{eq:umd}) without the correction term, within statistical uncertainties. Therefore, the two different momentum dependent interactions in Eq. (\ref{eq:umd}) and Eq. (\ref{umd1}) do not lead to significant differences in the observables.




\end{document}